\documentclass{article}
\usepackage{arxiv}
\usepackage[utf8]{inputenc} % allow utf-8 input
\usepackage[T1]{fontenc}    % use 8-bit T1 fonts
\usepackage[colorlinks=true, allcolors=blue]{hyperref}       % hyperlinks
\usepackage{url}            % simple URL typesetting
\usepackage{booktabs}       % professional-quality tables
\usepackage{amsfonts}       % blackboard math symbols
\usepackage{nicefrac}       % compact symbols for 1/2, etc.
\usepackage{microtype}      % microtypography
\usepackage{lipsum}		% Can be removed after putting your text content
\usepackage{graphicx}
\usepackage{natbib}
\usepackage{doi}
\usepackage{adjustbox}
\usepackage[version=4]{mhchem}

\title{ALMA detection of sulfur dioxide and carbon monoxide in the atmosphere of Neptune}

\author{ {\color{red}Arijit  Manna}\\%\thanks{Use footnote for providing further
	Midnapore City College\\
	Kuturia, Bhadutala, Paschim Medinipur, \\West Bengal, India 721129 \\
	\texttt{Mannaarijit@hotmail.com} \\
	\And
	{\color{red}Sabyasachi Pal} \\
	Indian Centre for Space Physics\\43 Chalantika, Garia Station Road, \\Kolkata, India 700084\\\\
	Midnapore City College\\
	Kuturia, Bhadutala, Paschim Medinipur, \\West Bengal, India 721129 \\
	\texttt{sabya.pal@gmail.com} \\
}

\renewcommand{\shorttitle}{\ce{SO2} and CO in Neptune}

\begin{document}
\maketitle

\begin{abstract}
	The space and ground-based observations have shown a lot of activities and instabilities in the atmosphere of the giant ice planet Neptune. Using the archival data of high resolution Atacama Large Millimeter/Submillimeter Array (ALMA) with band 7 observation, we present the spectroscopic detection of the rotational emission line of sulfur dioxide (\ce{SO2}) at frequency $\nu$ = 343.476 GHz with transition J=57$_{15,43}$--58$_{14,44}$. We also re-detect the emission line of carbon monoxide (CO) at frequency $\nu$ = 345.795 GHz with transition J=3--2. The molecular lines of \ce{SO2} and CO in the atmosphere of Nepure are detected with the $\geq$4$\sigma$ statistical significance. The statistical column density of \ce{SO2} is N(\ce{SO2}) = 2.61$\times$10$^{15}$ cm$^{-2}$ with rotational temperature $T_{\ce{SO2}}$ = 50 K and the statistical column density CO is N(CO) = 1.86$\times$10$^{19}$ cm$^{-2}$ with $T_{\ce{CO}}$ = 29 K. The typical mixing ratio in the atmosphere of Neptune for \ce{SO2} is 1.24$\times$10$^{-10}$ and CO is 0.88$\times$10$^{-6}$. The \ce{SO2} and CO gas in the atmosphere of Neptune may create due to Shoemaker-Levy 9 impacts in Jovian planets since the 1994.
\end{abstract}

% keywords can be removed

	\section{Introduction}
High abundant carbon monoxide (CO) gas is present in the atmosphere of Neptune \citep{ros92}. Recently, the emission line properties of CO indicated that a larger mole fraction of CO exists in the upper stratosphere of Neptune rather than the lower region \citep{lel00, hes07, fle10, lus13}. The heterogeneous vertical distribution leads to the downward transportation of CO which originates from an external source. To put it another way, the stratosphere of Neptune is more likely to come from somewhere else. The CO gas in the stratosphere of Neptune is most probably coming from comets or other candidates.

In general, the comet Shoemaker-Levy 9 (SL9) collisions with Jupiter in 1994 resulted in a large amount of CO, with a total mass similar to that found in the stratosphere of Neptune \citep{lel00}. Earlier, the cometary origin of CO has been proposed for both Saturn and Uranus \citep{cav10, cav14}. The mixing ratio of CO in the stratosphere of Uranus and Saturn was very smaller than the stratosphere of Neptune. These new findings may indicate that the stratospheres of gaseous planets have some extra species which originate from the cometary impact. Further research of the stratosphere of Neptune is essential for a deeper understanding of the cometary origin and its effect on the planetary atmosphere, as its atmosphere is likely to contain a greater amount of cometary originated gases than other planets.
\begin{table*}
	\caption{Details of ALMA observations and detected species in the atmosphere of Neptune.}
	\centering
		\begin{adjustbox}{width=0.98\textwidth}
	\begin{tabular}{|c|c|c|c|c|c|c|c|c|c|c|c|}
		\hline \hline
		Start date (UT)&Geocentric   &Angular&Spectral&Molecular& Frequency & Transition& $E_{L}$&$\Delta V$ \\
		yyyy-mm-dd     &distance (AU)&diameter (arcsec)&resolution (kHz)&Species& (GHz)&  (J)  & (K)&km s$^{-1}$\\
		\hline
		2016-04-30&30.468&2.230&1128.91&\ce{SO2}&343.476&57$_{15,43}$--58$_{14,44}$&2051.270&--42.68\\
		2016-04-30&30.468&2.230&1128.91&\ce{CO}&345.795&3--2&11.535&--590.70\\
		\hline
	\end{tabular}
\end{adjustbox}
	\label{tab:prop}
\end{table*}
\begin{figure*}
	\centering
	\label{fig:so}
	\includegraphics[width=0.48\textwidth]{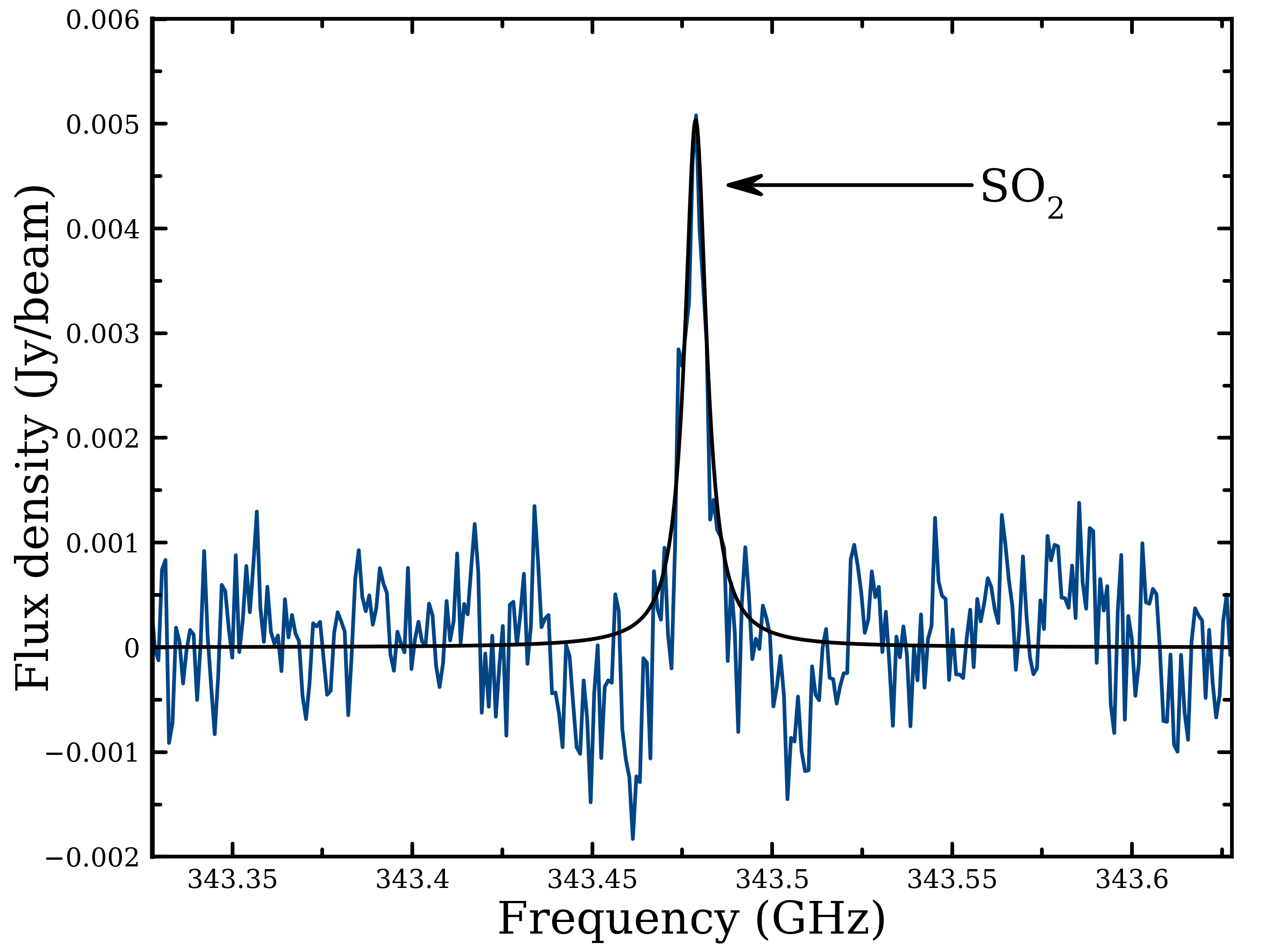}	\includegraphics[width=0.48\textwidth]{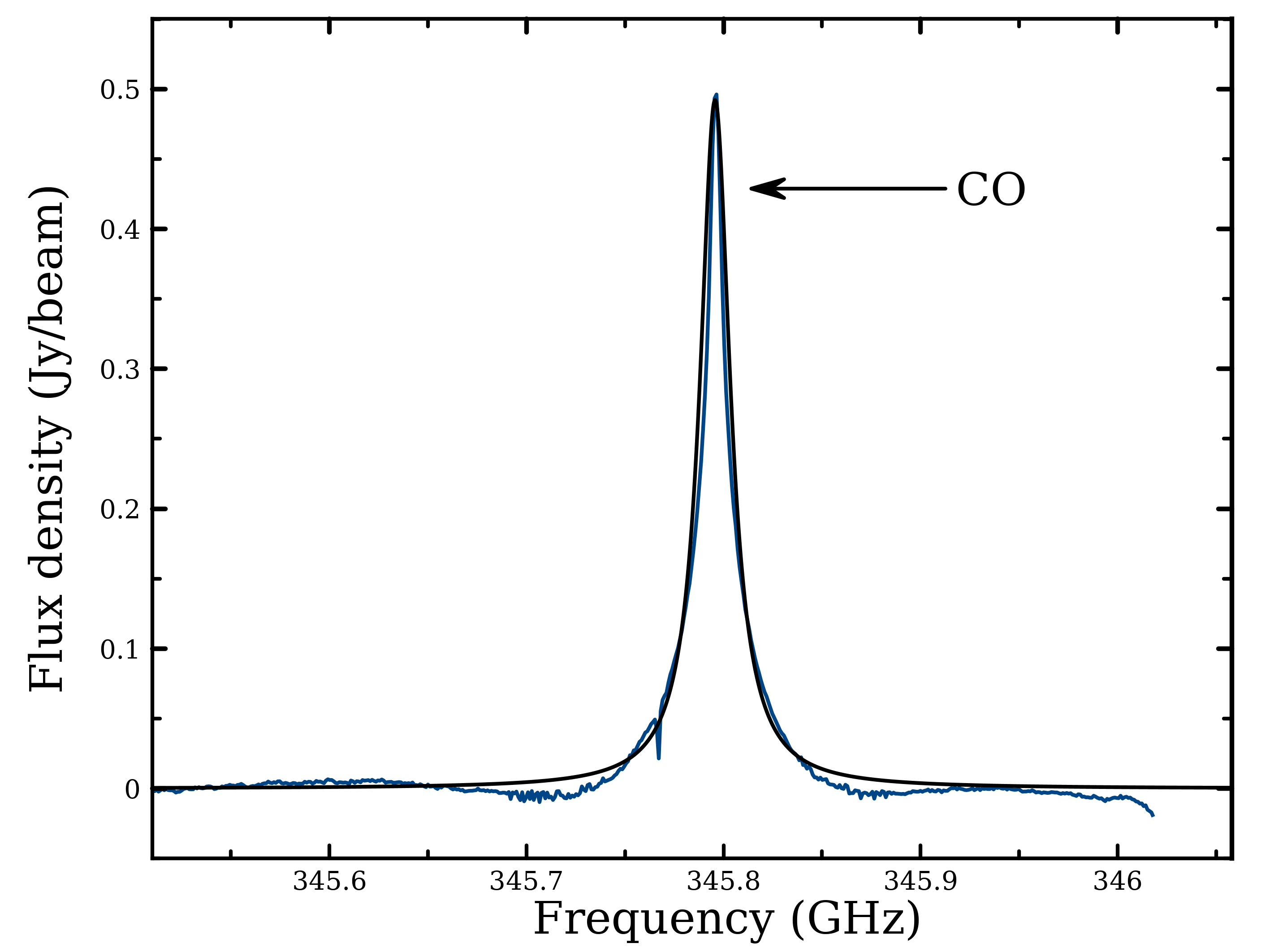}
	\caption{Rotational emission spectrum of \ce{SO2} and CO in the atmosphere of Neptune at the frequency of $\nu$ = 343.476 and 345.795 GHz with transition J=57$_{15,43}$--58$_{14,44}$ and J=4--3 using ALMA band 7 observation.  The spectral coverage in the figure is 250 MHz for \ce{SO2}line and 400 MHz for CO line with a total integration time of 876.960 second. In the emission spectrum, the solid black line corresponds to the best fit model of column density.}
\end{figure*}

Different varieties of organisms are imported into the planetary atmosphere after the deposition of vaporized cometary nuclei in the atmosphere due to a large cometary impact. In the Jovian stratosphere, \ce{S2}, \ce{CS2}, OCS, CO, \ce{H2S}, HCN, CS and \ce{H2O} species were detected after the SL9 event \citep{nol95,lel95,spr96,mor01}. These SL9 species may produce in the stratosphere of Neptune. We focus on the sulfur-bearing species in the stratosphere of Neptune. Both S and O atoms in the emitted gases were primarily supplied by the SL9 incident \citep{mor03}. The discoveries of S-bearing species with limited distribution in Neptune's CO-rich area will support not only the existence of a recent cometary impact but also provide new insights into the sulfur chemistry caused by the cometary impact. Earlier, \cite{ino14} did not detect any sulfur-bearing species including \ce{SO2} at $\nu$ = 338.612 GHz and CS at $\nu$ = 342.882 GHz in the stratosphere of Neptune using Atacama Sub-millimeter Telescope Experiment 10-m single-dish telescope and later \cite{mor17} did present the detection of CS in Neptune using ALMA at 342.882 GHz. We successfully detect the emission line of \ce{SO2} using ALMA at 343.476 GHz.

In this article, we present the first spectroscopic detections of \ce{SO2} at $\nu$ = 343.476 GHz and re-detection of CO at $\nu$ = 345.785 GHz in the atmosphere of Neptune using ALMA band 7 data. In Sect.~\ref{obs}, we discuss the observations and data reductions. The result and discussion of the detection of \ce{SO2} and CO is shown in Sect.~\ref{res}. The summary is presented in Sect.~\ref{sec:discussion}.

\section{Observations and data reduction}
\label{obs}

The high-resolution interferometric study of Neptune was observed on 30-Apr-2016 between 11:09:22.1 to 11:48:47.0 UTC which is a part of the ALMA project {\tt 2015.1.01471.S}. The observation was done with the 12 m antennae in the telescope array by using the band 7 dual sideband receiver. Forty-one antennas were working during the observation. The correlator was set up to four basebands with frequencies 354.22--356.09 GHz and 356.03--357.90 GHz in upper sideband and 342.46--344.33 GHz and 344.14--346.02 GHz in the lower sideband. The observation details are shown in Tab.~\ref{tab:prop}. During the observation, the weather condition was very good with precipitable water vapor (PWV) 1.38 mm. The source Pallas was used as a flux calibrator, J0006--0623 was used as a bandpass calibrator and, J2246--1206 was used as a phase calibrator. The telescope was set up to monitor the ephemeris location of Neptune and update the phase center coordinates in real-time.

The raw data of Neptune in {\tt ASDM} format was taken from the ALMA\footnote{\href{https://almascience.nao.ac.jp/}{https://almascience.nao.ac.jp/}} Science Archive. During the data analysis procedure, we applied routine flagging, bandpass calibration, and complex gain calibration. For each baseline, the calculated continuum flux density was scaled to fit the Butler-JPL-Horizons 2012 \citep{But12} Pallas flux model, which is estimated to be appropriate to within 15\%. The {\tt uvcontsub} task in the {\tt CASA (version 4.2.1)}\footnote{\href{https://casa.nrao.edu/}{https://casa.nrao.edu/}} was used for continuum subtraction of the visibility amplitudes, and the {\tt tclean} task was used for imaging. We applied the Hogbom algorithm to deconvolve the point-spread function (PSF) for each spectral channel, with natural visibility weighting and having a threshold flux level of twice the predicted RMS noise. The resultant beam of the emission map of \ce{SO2} and CO is 0.406$^{\prime\prime}\times$0.350$^{\prime\prime}$ and 0.406$^{\prime\prime}\times$0.352$^{\prime\prime}$ respectively. Using the JPL Horizons topocentric radial velocity, the spectral coordinate scale of each image was Doppler-shifted to rest frame of Neptune and the images were converted to (projected) linear distances from equatorial coordinates with respect to the center of Neptune.

\section{Results and discussions}
\label{res}
\subsection{Emission line of \ce{SO2} and CO in the atmosphere of Neptune}
In the atmosphere of Neptune, we discover the solid rotational emission line of \ce{SO2} and CO at frequency $\nu$ = 343.476 and 345.795 GHz respectively. The sulfur dioxide and carbon monoxide emission spectrum is created by integrating the reduced ALMA data cubes within a circular area of radius 1.2$^{\prime\prime}$ from the center of Neptune. Fig.~\ref{fig:so} displays the molecular rotational emission spectra of \ce{SO2} and CO in the atmosphere of Neptune. Tab.~\ref{tab:prop} summarises the molecular properties of detected \ce{SO2} and CO species. The online Splatalogue\footnote{\href{https://splatalogue.online//}{https://splatalogue.online//}} database for astronomical molecular spectroscopy is used to confirm the spectral peak of \ce{SO2} and CO emission lines.

\subsection{Radiative transfer modeling of \ce{SO2} and CO}
For radiative transfer simulation of the emission lines of \ce{SO2} and CO, we use the {\tt XCLASS}\footnote{\href{https://xclass.astro.uni-koeln.de/}{https://xclass.astro.uni-koeln.de/}} package in {\tt CASA}. The molecular properties of observed molecules in the atmosphere of Neptune are taken from the Jet Propulsion Laboratory \citep{pic98} and CDMS \citep{Mu01} spectroscopic databases. Using the aggregated values of the partition function between 9.375 and 300 K, which are available in the version of {\tt XCLASS}, a global linear fit in log-log space is obtained for each species. The detected species is described as a vibrationally excited state of a molecule, and the modeling was done on a species-by-species basis. We assume that the population levels of each species are represented by a single excitation temperature ($T_{rot}$) during the radiative transfer simulation using the {\tt XCLASS}. This argument holds true at both low and high densities, where $T_{rot}$ equals the temperature of such cosmic microwave background, $T_{CMB}$ = 2.73 K, and even where collisions are regular enoug to support the Local Thermodynamic Equilibrium (LTE) approximation.

We measure the statistical column density of \ce{SO2} and CO and compare them with the \ce{H2}, to extract the abundance of observed species in the atmosphere of Neptune. The beam average column density of \ce{SO2} is N(\ce{SO2}) = 2.61$\times$10$^{15}$ cm$^{-2}$ with rotational temperature $T_{\ce{SO2}}$ = 50 K and the beam average column density of CO is N(CO) = 1.86$\times$10$^{19}$ cm$^{-2}$ with $T_{\ce{CO}}$ = 29 K. The typical mixing ratios of \ce{SO2} and CO, in the atmosphere of Neptune, are 1.24$\times$10$^{-10}$ and 0.88$\times$10$^{-6}$ respectively.

\section{Summary}
\label{sec:discussion} 

Using the ALMA band 7 observation, we have confirmly detect the presence of the rotational emission line of \ce{SO2} and CO in the atmosphere of Neptune at frequency $\nu$ = 343.476 and 345.795 GHz with $\geq$4$\sigma$ statistical significance. The abundance of \ce{SO2} and CO in the atmosphere of Neptune is 1.24$\times$10$^{-10}$ and 0.88$\times$10$^{-6}$ respectively. Earlier, using ASTE single-dish telescope, \cite{ino14} did not detect any sulfur species including \ce{SO2} and they found the upper limit of \ce{SO2} in the atmosphere of Neptune is 9.2$\times$10$^{-10}$ which occurred due to SL9 events. Our calculated mixing ratio of CO in the atmosphere of Neptune is similar to \cite{mor17}. Further spectroscopic measurements of the dynamic molecular gases in the atmosphere of Neptune can help to know the origin and formation process of detected trace gases which originated probably during the SL9 events.

\section*{Acknowledgement}
This paper makes use of the following ALMA data: ADS /JAO.ALMA\#2015.1.01471.S. ALMA is a partnership of ESO (representing its member states), NSF (USA), and NINS (Japan), together with NRC (Canada), MOST and ASIAA (Taiwan), and KASI (Republic of Korea), in co-operation with the Republic of Chile. The Joint ALMA Observatory is operated by ESO, AUI/NRAO, and NAOJ. The data that support the plots within this paper and other findings of this study are available from the corresponding author upon reasonable request. The raw ALMA data are publicly available at \href{https://almascience.nao.ac.jp/asax/}{https://almascience.nao.ac.jp/asax/}.\\

\section*{Data Availability Statement}
The data that support the plots within this paper and other findings of this study are available from the corresponding author upon reasonable request. The raw ALMA data are publicly available at \\\href{https://almascience.nao.ac.jp/asax/}{https://almascience.nao.ac.jp/asax/} (project id: 2015.1.01471.S).

\end{document}